# Tip induced unconventional superconductivity on Weyl semimetal TaAs

He Wang[1,2,†], Huichao Wang[1,2,†], Yuqin Chen[1,2,†], Jiawei Luo[1,2], Zhujun Yuan[1,2], Jun Liu[3], Yong Wang[3], Shuang Jia[1,2], Xiong-Jun Liu[1,2*], Jian Wei[1,2*], and Jian Wang[1,2*]

[1]International Center for Quantum Materials, School of Physics, Peking University, Beijing 100871, China

[2]Collaborative Innovation Center of Quantum Matter, Beijing, China

[3]Center of Electron Microscopy, State Key Laboratory of Silicon Materials, Department of Materials Science and Engineering, Zhejiang University, Hangzhou 310027, China

[†]These authors contributed equally to this work.

*e-mail: jianwangphysics@pku.edu.cn; weijian6791@pku.edu.cn; xiongjunliu@pku.edu.cn

**Weyl fermion is a massless Dirac fermion with definite chirality, which has been long pursued since 1929. Though it has not been observed as a fundamental particle in nature, Weyl fermion can be realized as low-energy excitation around Weyl point in Weyl semimetal[1-10], which possesses Weyl fermion cones in the bulk and nontrivial Fermi arc states on the surface. As a firstly discovered Weyl semimetal, TaAs crystal[4-10] possesses 12 pairs of Weyl points in the momentum space, which are topologically protected against small perturbations. Here, we report for the first time the tip induced superconductivity on TaAs crystal by point contact spectroscopy. A conductance plateau and sharp double dips are observed in the point contact spectra, indicating p-wave like unconventional superconductivity. Furthermore, the zero bias conductance peak in low temperature regime is detected, suggesting potentially the existence of Majorana zero modes. The experimentally observed tunneling spectra can be interpreted with a novel mirror-symmetry protected topological superconductor induced in TaAs, which can exhibit zero bias and double finite bias peaks, and double conductance dips in the measurements. Our**

**work can open a broad avenue in search for new topological superconducting phases from topological Weyl materials and trigger intensive investigations for pursuing Majorana fermions.**

Extensive investigations have been conducted for observing topological superconductivity and the long sought after Majorana fermions[11-17], which have potential applications to fault-tolerant topological quantum computation[18,19]. Especially, the progress of topological materials has stimulated the extensive studies in this direction[14-17]. Most previous proposals and experiments are based on inducing topological superconductivity in spin-orbit coupled systems by proximity to conventional superconductors[20-25]. Alternatively, by using hard point contact on topological Dirac semimetal[26,27], unconventional superconductivity[28,29] may have been induced and detected. Very recently, the Weyl semimetal was remarkably discovered in the TaAs crystal, which has gapless nodal points in the bulk and Fermi arc states in the boundary. Compared with the Dirac semimetal, in a Weyl semimetal the bulk nodal points have definite chirality and are protected by topology[1-10]. This essential difference may enable the Weyl semimetal to be a more natural candidate to realize a topological superconductor if superconductivity can be induced.

TaAs single crystals in our work were grown by chemical vapor transport method[10]. The atomically high-resolution transmission electron microscopy image of one typical sample is shown in Fig. 1a, which manifests a high-quality nature. Transport data are mainly obtained from sample 1 (s1). Its temperature dependence of bulk resistivity down to 2 K measured by standard four-electrode configuration demonstrates a metallic-like behavior (Fig. 1b). The observation of chiral anomaly induced negative magnetoresistance gives the crucial transport signature for Weyl fermions in Weyl semimetal phase (lower inset of Fig. 1b)[9,10]. In a perpendicular field (B//$c$ axis), the longitudinal magnetoresistance of s1 exhibits obvious Shubnikov–de Haas (SdH) oscillations below 80 K (Fig. 1c). Fig. 1d shows the SdH oscillation at 2 K after subtracting a polynomial background from the original data. A major frequency at F=8 T is obtained through Fourier transformation. The second harmonics 2F=16 T, likely due to spin-splitting, is also observed. Lower inset of Fig. 1d shows the Landau index n with respect to 1/B. The linear fitting intercepts around zero, suggesting the π Berry's phase of TaAs crystal associated with Weyl fermions. The above results indicate that our TaAs crystal shows typical electrical transport properties of Weyl semimetal.

Point contact (PC) measurements are further conducted in the "needle-anvil configuration" by approaching the (001) surface of s1 to a mechanically sharpened PtIr tip (inset of Fig. 2a). Differential resistance (d$V$/d$I$) is measured by standard lock-in technique in quasi-four-probe configuration, and all the point contact spectra shown in the text are normalized to normal state conductance. Fig. 2a shows the temperature dependence of the PC resistance with normal state resistance of 18.8 Ω. Interestingly, a significant resistance drop is observed in the zero-field cooling (ZFC) process, suggesting existence of tip-induced superconductivity in the local regime of PtIr-contacted TaAs crystal[30,31]. Furthermore, the onset temperature ($T_c^{onset}$) of the resistance drop decreases from 5.9 K to 4.75 K when a perpendicular filed of 3 T is applied, consistent with the existence of superconductivity.

To further study the observed superconductivity in Weyl semimetal TaAs, the point contact spectra (PCS) at 0.5 K are obtained at selected perpendicular magnetic fields (Fig. 2b). The PCS at zero magnetic field is shown by the navy blue curve, showing two features: a small zero bias conductance peak (ZBCP) and a wide conductance plateau ended with sharp dips symmetric to zero bias. Such features cannot be fitted by conventional BTK model for superconducting order parameter with s-wave symmetry. For comparison, PCS of PtIr tip on a Ta foil can be well fitted by modified 1D BTK model with a gap value of 0.26 mV at 1.2 K (see Fig. S1a in Supplementary Information). As the magnetic field increases, the conductance plateau shrinks, the "wiggles" beyond the conductance dips diminish and a multiple step-like feature is finally formed at 3 T, possibly due to removal of the degeneracy of order parameter symmetry. When field is applied parallel to the (001) surface of s1, there is little influence on the PCS, indicating strong anisotropy of the order parameter (see Fig. S1b in Supplementary Information).

The PCS of s1 at selected temperatures from 0.5 K to 6 K are presented in Fig. 2c. Different from the field dependence of PCS there is no step-like feature with increasing temperature as obtained with increasing field. The ZBCP in PCS can be interpreted by several mechanisms[32], the relevant one here is related to the Majorana zero energy mode. In this scenario, ZBCP broadens with increasing temperature and should not split when external magnetic field is applied. The observed ZBCP at different temperatures are shown in Fig. 2d. With increasing temperature, it becomes broader and finally invisible at 2.5 K, which is further illustrated by the temperature dependence of the estimated full width at half maximum (FWHM) of ZBCP (see Fig. S1d in Supplementary Information). Under external magnetic field, no splitting of ZBCP is observed either (Fig. 2b). Thus, the above experimental observations are consistent with the picture of Majorana zero energy mode.

The property of the tip-induced superconductivity is further investigated as the normal state resistance of this PC is varied from 18.8 Ω to 4.2 Ω by pushing the sample towards the tip. The PCS at 0.6 K is shown by the darkest blue curve in Fig. 3a. Similar features such as the conductance plateau and ZBCP (Fig. 3b) are observed. The conductance dip near the edge of the plateau locates at 1.1 mV, close to that of the PC state of 18.8 Ω at 0.5 K (Fig. 2b,c), suggesting this is a spectroscopic energy scale of the superconducting gap. Such conductance dips in PCS can be attributed to chiral/helical p-wave superconductivity or critical current effect. The critical current mechanism is less probable in this situation (see Sec III in supplementary information). For the p-wave situation, the gap energy is located around the conductance dip[33], and the temperature dependence of gap amplitudes for PC states of 18.8 Ω and 4.2 Ω deviates the standard BCS dependence (see Supplementary Fig. S1c). The PCS at different perpendicular magnetic fields are shown in Fig. 3b. In the inset of Fig. 3b, the $T_c^{onset}$ at zero magnetic field is 5.6 K, which is suppressed to 4.4 K by applying a perpendicular field of 3 T. Both Fig. 2a and inset of Fig. 3b suggest that the zero temperature upper critical magnetic field of the induced superconductivity is much higher than 3 T, which is large compared to the gap value and indicates the observation of unconventional superconductivity. Additionally, a tungsten tip is also used to make point contact on TaAs single crystal and superconductivity is also induced (see Fig. S2 in Supplementary Information).

The observed tunneling spectra suggest the existence of Majorana surface states associated with the superconductivity induced by tip. Here we propose that the experimental observations can be interpreted with a minimal model of topological superconductor obtained in TaAs, with the nontrivial topology of the induced superconductivity being partially inherited from the parent topological Weyl semimetal material (Supplementary Information, Sec IV). The TaAs has 12 pairs of Weyl nodes which are sketched in Fig. 4f. These nodal points are related by mirror and C4 symmetries. Note that the superconductivity is induced on the (001) surface, which preserves the both mirror symmetries $M_x$ and $M_y$, with $M_i = s_i$ being Pauli matrices acting on real spin space. We then seek for the nontrivial superconducting states respecting such mirror symmetries[34].

For convenience, we consider the superconducting state induced on the Weyl cones A, B, C, and D (Fig. 4f), of which the Weyl cones A and B are time reversal partners of C and D, respectively. The superconductivity on other Weyl cones can be obtained via C4 transformation or studied in the similar way. In the presence of time-reversal symmetry, a uniform superconducting pairing must occur between Weyl cones A (B) and C (D). Furthermore, by some analysis we find that to fully gap out the bulk band, the possible pairing terms respecting both time-reversal and mirror symmetries include $\mathcal{H}_{\Delta_s} = \Delta_s s_y \tau_y$, $\mathcal{H}_{\Delta_{px}} = \Delta_{px}(k_x) s_y \tau_x$, and $\mathcal{H}_{\Delta_{py}} = \Delta_{py}(k_y) \tau_y$, where $\Delta_s$ is a constant while $\Delta_{pi}(-k_i) = -\Delta_{py}(k_i)$ has odd parity, with $\tau_{x,y}$ being Pauli matrices defined in the Nambu space (Supplementary Information, Sec IV). and $\mathcal{H}_{\Delta_{py}} = \Delta_{py}(k_y) \tau_y$, with $\Delta_{pi}(-k_i) = -\Delta_{pi}(k_i)$. Note that in the presence of superconductivity, the mirror symmetry $M_y$ has a second choice as $M_y = s_y \tau_z$. The mirror invariant planes for the mirror symmetry $M_i$ are $k_i = 0, \pi$. A 3D mirror topological state can be characterized by the mirror topological invariant defined by[35]

$$N_{M_i} = sgn[C_1(k_i = 0) - C_1(k_i = \pi)][|C_1(k_i = 0)| - |C_1(k_i = \pi)|],$$

with $C_1(k_i = 0, \pi)$ being the mirror Chern number for the 2D mirror plane $k_i = 0, \pi$. A straightforward calculation shows that $N_{M_i} = 0$ for the pairing order $\mathcal{H}_{\Delta_s}$ and $\mathcal{H}_{\Delta_{px}}$, while $N_{M_y} = 2$ for the superconducting order $\mathcal{H}_{\Delta_{py}}$. We then conclude that with odd parity superconducting order $\mathcal{H}_{\Delta_{py}}$, the Weyl semimetal TaAs becomes a 3D fully gapped mirror topological superconductor, and the mirror invariant $N_{M_i}$ describes the number of Majorana surface Dirac cones existing on the boundary[35].

The numerical results are shown in Fig. 4a-e by taking the superconducting order as $\Delta_{py} = \Delta_t \sin(k_y)$. The energy spectra are shown in Fig. 4a,b, with $E_g = |\Delta_{py}(k_f)|$ being the bulk superconducting gap and $k_f$ the Fermi momentum. The surface state dispersion along $x$ direction is inherited from that of the original Fermi arc states along this direction, while the dispersion along $y$ direction is governed by the superconducting order. This renders the highly anisotropic Majorana surface Dirac cones, with the momentum-resolved local density of states being plotted in Fig. 4c, d, where a peak is clearly identified at zero energy in each case. The tunneling spectra due to the Majorana surface states are shown in Fig. 4e. It can be seen that in the regime with a relatively small tunneling energy $\delta$, a ZBCP, together with finite bias double conductance peaks and double conductance dips are obtained with the present model. These numerical results are consistent with the experimental observations.

In summary, a superconducting phase with the critical temperature up to 5.9 K has been induced at the interface between a hard normal metal tip and Weyl semimetal TaAs single crystal. The detection of ZBCP and double conductance dips indicate the existence of Majorana zero energy modes and unconventional superconductivity. Our further theoretical analyses suggest that a novel mirror-symmetry protected topological superconductor could be induced on TaAs, which is consistent with the experimental observations.

## Acknowledgements

We acknowledge Hong Lu and Yanan Li for the help in experiments. This work was financially supported by National Basic Research Program of China (Grant Nos. 2013CB934600, 2012CB927400, and 2012CB921300), the Research Fund for the Doctoral Program of Higher Education (RFDP) of China，the Open Project Program of the Pulsed High Magnetic Field Facility (Grant No. PHMFF2015002), Huazhong University of Science and Technology, Open Research Fund Program of the State Key Laboratory of Low-Dimensional Quantum Physics.

## Author Contributions

J.Wang and J.Wei conceived and instructed the experiments. He Wang, Huichao Wang and J.W.L. carried out transport measurements. Y.Q.C. and X.J.L. proposed the theoretical model to interpret experimental observation. Z.J.Y. and S.J. grew the crystals. J.L. and Y.W. made the TEM study.

## Additional information

Supplementary information is available in the online version of the paper. Reprints and permissions information is available online at www.nature.com/reprints.
Correspondence and requests for materials should be addressed to J.Wang, J.Wei and X.J.L.

## Competing Financial Interests statement

The authors declare no competing financial interests.

# Figure Legends

**Figure 1| Characterization and electrical transport properties of TaAs. a**, High-resolution transmission electron microscopy image of the TaAs single crystal. **b**, Temperature dependence of four-probe bulk resistivity of sample 1 (s1) showing typical metallic behavior with a residual resistance at low temperatures. Upper Inset: Schematics of the standard four-probe method measurement configuration. Lower inset: Negative magnetoresistance as a signature of chiral anomaly in s1 when B//E. **c**, Normalized longitudinal magnetoresistance of s1 with obvious Shubnikov–de Haas (SdH) oscillations at selected temperatures in the perpendicular magnetic field up to 15 T. **d**, SdH oscillations of s1 after subtracting the background from raw data at 2 K in **c**. Upper Inset: Fourier transformation of the oscillations showing peaks at F=8 T and 2F=16 T. Lower inset: Linear fitting of the Landau fan diagram indicating a non-trivial $\pi$ Berry's phase of TaAs.

**Figure 2| Point contact measurements between a PtIr tip and the TaAs crystal (s1). a**, Temperature dependence of zero-bias PC resistance showing typical superconducting transition. Inset: Schematics of the PCS measurement configuration. **b,** Normalized d$I$/d$V$ spectra at 0.5 K under different out-of-plane magnetic fields. The navy blue line denotes zero field measurement. **c,** Normalized d$I$/d$V$ spectra at selected temperatures from 0.5 to 6.0 K. **d,** Zoom in of the normalized ZBCP.

**Figure 3| Point contact measurements in another PC state of 4.2 Ω. a,** Normalized d$I$/d$V$ spectra for the temperature range from 0.6 to 6.0 K. Inset: Zoom-in of the normalized ZBCP at 0.6

K. **b**, Normalized d$I$/d$V$ spectra at 4.5 K under different out-of-plane magnetic fields. Inset: Temperature dependence of the PC resistance in ZFC or field cooling (FC) process (H$_\perp$=3 T).

**Figure 4| Theoretical interpretation of the experimental observation. a-b,** Energy spectra of the bulk and surface states in the presence of an odd-parity superconductivity on the Weyl cones A-D sketched in (f), which respects both time-reversal and $M_{x,y}$ mirror symmetries. $E_0$ relates to the energy of Van Hove singularity point, and $E_g$ is the superconducting gap. **c-d**, Momentum resolved local density of states (LDOS). A peak is obtained at the zero energy. **e**, The tunneling spectra at temperature much lower than the superconducting gap. A ZBCP, together with finite bias double conductance peaks and double conductance dips are obtained for the relatively small tunneling energy $\delta = 0.01 E_g$ (red curve). The double conductance peaks and dips can still be observed with a relatively large tunneling energy (blue curve). **f**, The sketch of the Weyl points in TaAs.

**Figure 1**

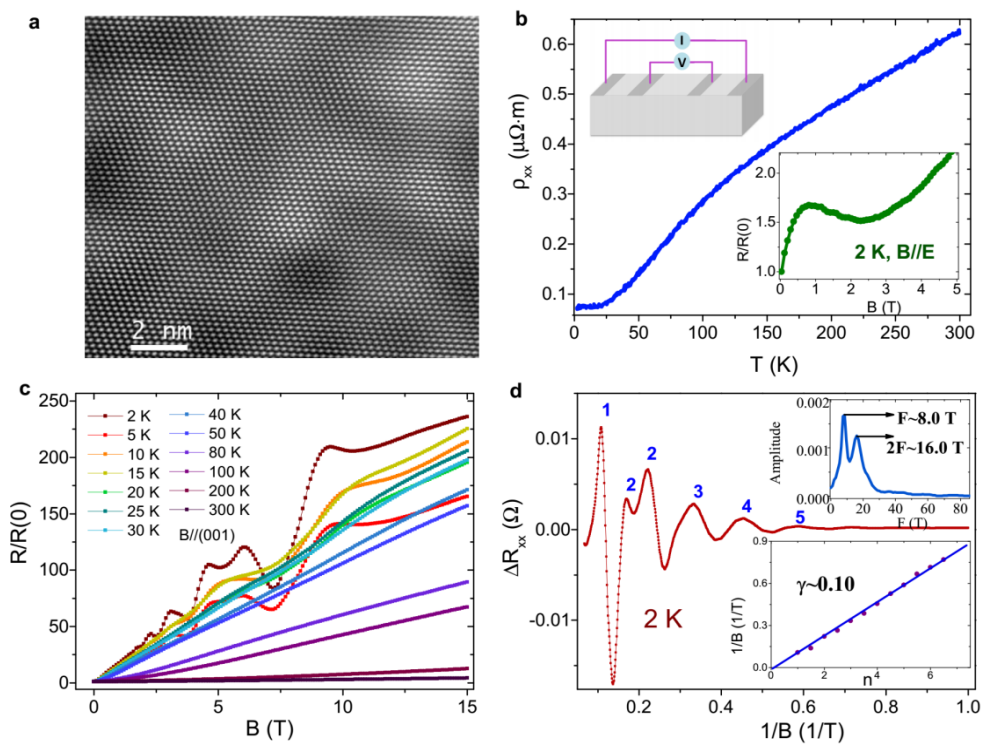

**Figure 2**

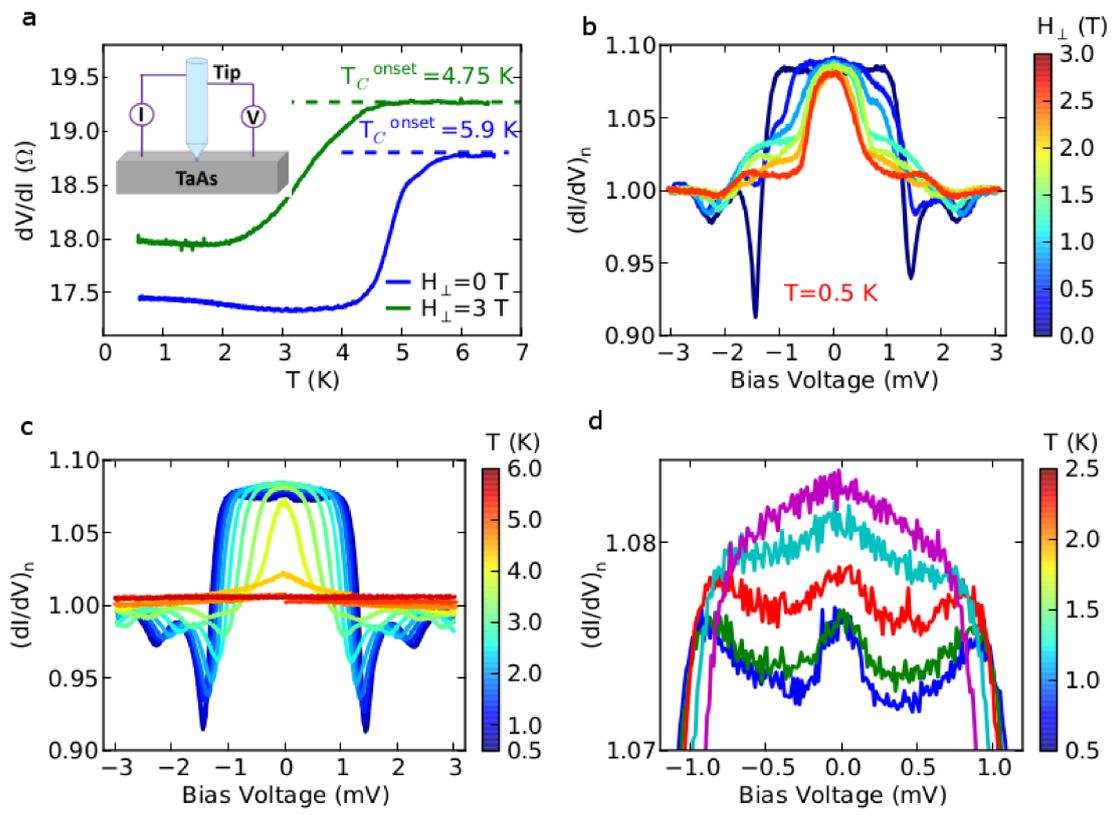

**Figure 3**

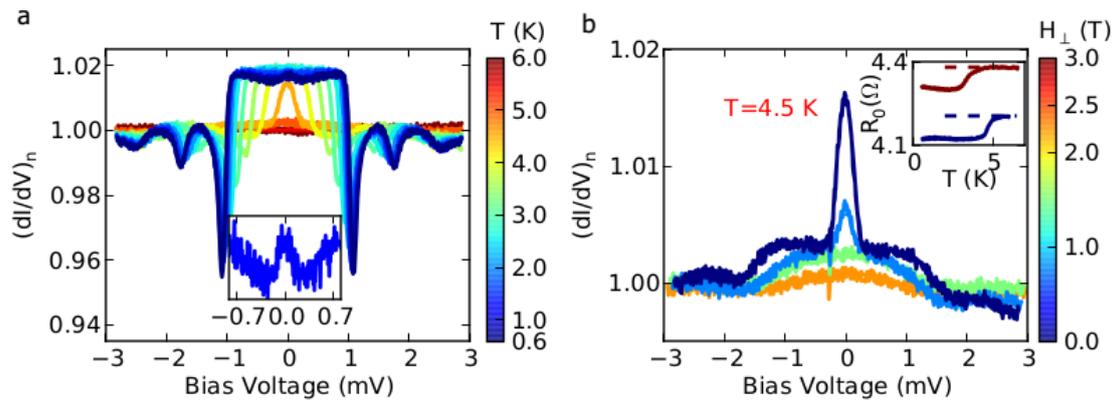

**Figure 4**

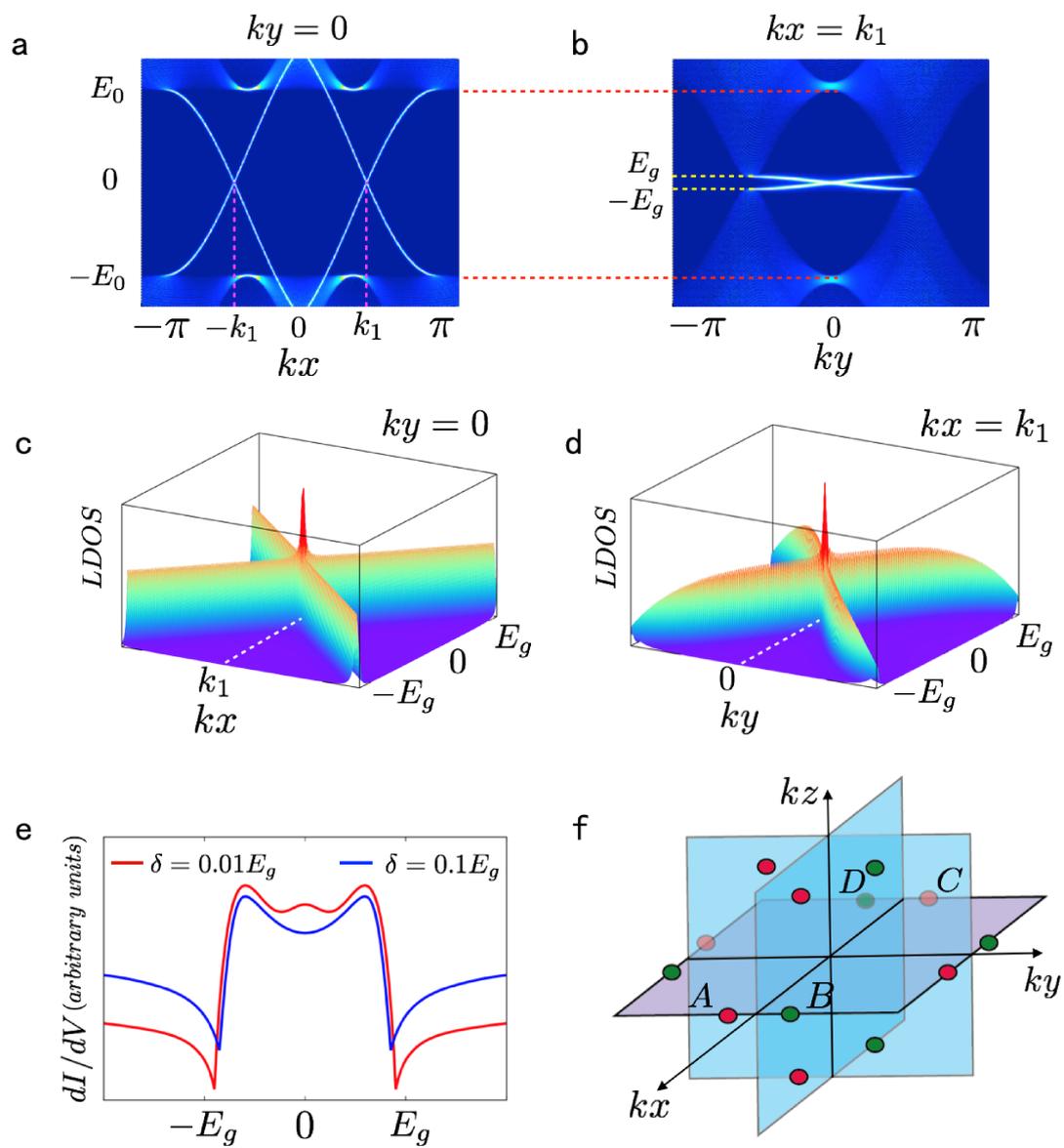

# Methods

High-resolution transmission electron microscopy image of TaAs crystal was obtained by the FEI Tecnai F20 transmission electron microscope operated at 200 kV. And the electrical transport measurements by standard four-probe method were conducted in a Physics Property Measurement System (PPMS-16T). Electrical contacts were made by using 25-μm-diameter Au wires and silver paint.

Point contact measurements were carried out between a mechanically sharpened PtIr tip and the (001) surface of TaAs single crystal mounted on an Attocube nanopositioner stack. The low temperature environment were obtained by using a dry dilution refrigerator from Leiden, in which three-axis superconducting vector magnets were mounted with the maximum field of 1/1/3 Tesla in X/Y/Z direction. To ensure a stable point contact between the tip and sample, suspending spring system was applied at the bottom of the probe for vibration damping.

Supplementary Information

# Tip induced superconductivity on Weyl semimetal TaAs

He Wang[1,2,†], Huichao Wang[1,2,†], Yuqin Chen[1,2,†], Jiawei Luo[1,2], Zhujun Yuan[1,2], Shuang Jia[1,2], Xiong-Jun Liu[1,2,*], Jian Wei[1,2,*], and Jian Wang[1,2,*]

[1]*International Center for Quantum Materials, School of Physics, Peking University, Beijing 100871, China*

[2]*Collaborative Innovation Center of Quantum Matter, Beijing, China*

[†]These authors contributed equally to this work.

*e-mail: jianwangphysics@pku.edu.cn; weijian6791@pku.edu.cn; xiongjunliu@pku.edu.cn

# I Additional point contact measurements and analysis.

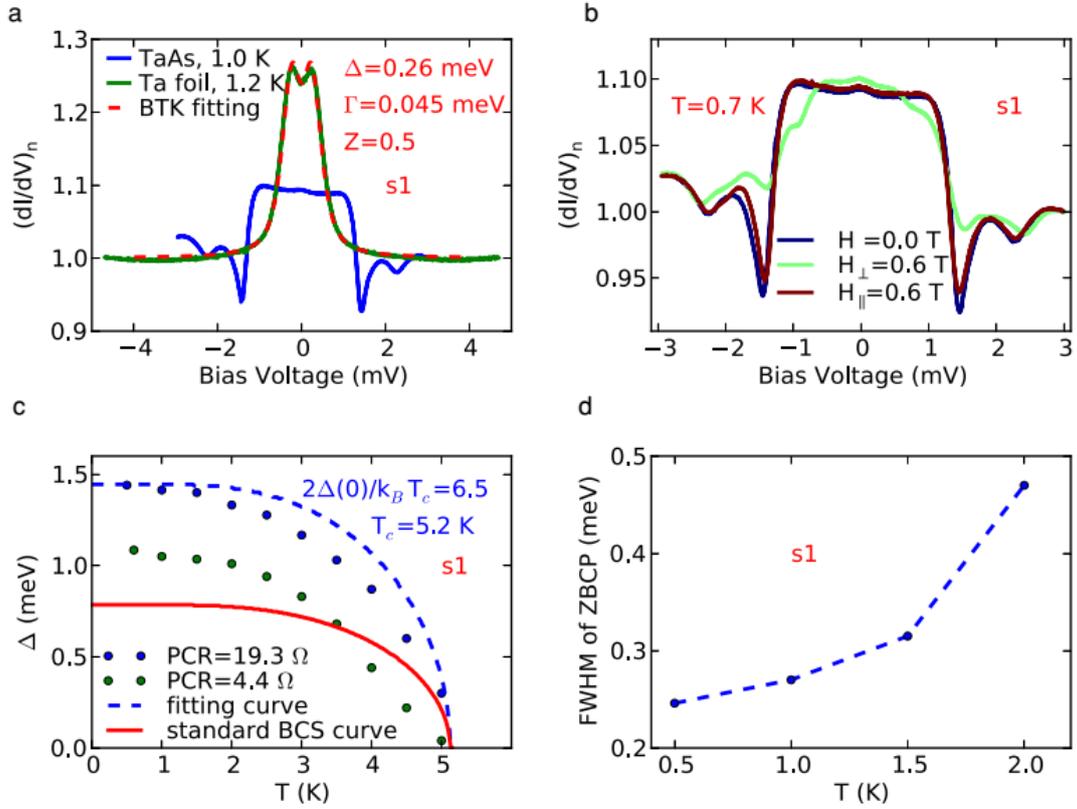

**Figure S1| Additional point contact spectra (PCS) and analyses. a,** Comparison between the PCS of PtIr tip on a Ta foil and a TaAs crystal. The modified 1-D BTK model can be applied on the PCS of Ta, indicating a gap value of 0.26 meV at 1.2 K. **b,** PCS under in-plane and out-of-plane magnetic fields showing anisotropic property of the tip induced superconductivity on a TaAs crystal at T=0.7 K. **c,** Temperature dependence of the gap determined by the conductance dips in Figs. 2 and 3 deviating from the standard BCS predication. **d,** FWHM (full width at half maximum) of ZBCP in Fig. 2d becomes smaller with decreasing T.

## II the point contact measurements with a tungsten tip

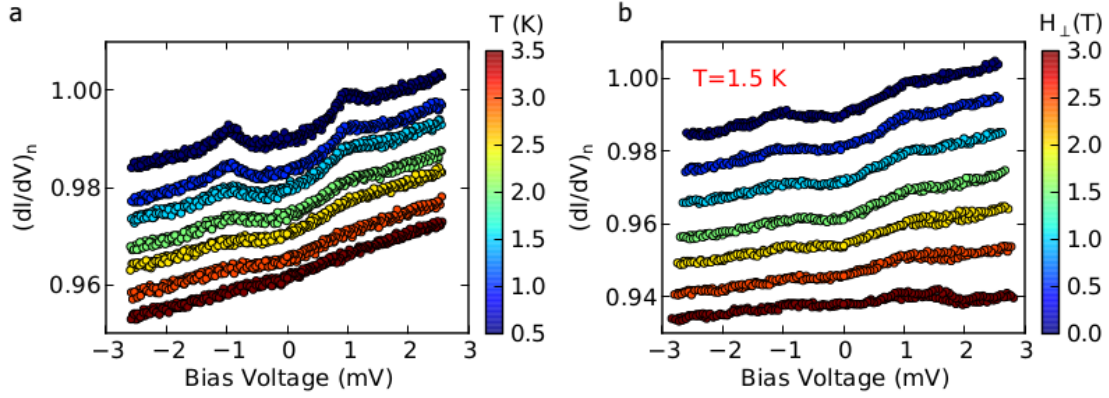

**Figure S2| PCS with a tungsten tip on the (001) surface of s1. a,** PCS at different temperatures from 0.5 to 3.5 K. **b,** PCS at 1.5 K under different out-of-plane magnetic fields (B//(001) surface). Proper shift has been done for clarity.

## III Discussion of alternative explanation of conductance dips

The conductance dips in PCS may be induced by chiral/helical p-wave superconductivity[1] or critical current effect. In the latter case, the Maxwell resistance ($R_m$) is comparable to the Sharvin resistance ($R_{sh}$)[2,3], as explained in the following.

The traditional way to justify a PC in ballistic regime is to compare the PC radius with mean free path. With the assumption of no barrier between the tip and the sample (Z=0 in the BTK model[4]), the PC radius can be estimated by the Sharvin formula. In practice, such simple model doesn't apply to most PtIr/TaAs PC states with finite barrier (Z >0). Considering the property of TaAs, one method is proposed here to justify that the PC states are close to the ballistic limit by analyzing the contribution of $R_m$ to the total PC resistance.

One method is to calculate the ratio between the Maxwell resistance and the Sharvin resistance. For a general hetero-PC, its resistance can be simply written as the sum of Sharvin resistance and the Maxwell resistance (reference)[5]:

$$R_{PC} = R_{sh} + R_m. \qquad (1)$$

$R_{sh}$ depends only on the contact radium in free-electron approximation in the Z=0 limit, which can be shown as

$$R_{sh} = \frac{2R_q}{k_F^2 a^2}, \qquad (2)$$

where $R_q = h/e^2$ is quantum resistance with the value of 25.8 kΩ, $a$ is the radius of the PC. $k_F$ is Fermi vector. In the ballistic limit, the $R_{PC}=R_{sh}$.

The $R_m$ can be written as:

$$R_m = \frac{\rho_{sample}}{2a} + \frac{\rho_{tip}}{2a} = R_{m-sample} + R_{m-tip}, \qquad (3)$$

where $\rho_{sample}$ ($\rho_{tip}$) is the bulk resistivity of the sample (tip). For a normal metal/superconductor PC out of the ballistic regime ($R_m$ is finite), when the bias current density exceeds the critical current density of the superconductor under the tip, a conductance dip may appear in PCS due to sudden change of $R_{m\text{-}sample}$ from zero to a finite value[2,3]. So the critical current effect would induce an observable conductance dip when $R_m$ is comparable to $R_{sh}$.

For our PtIr/TaAs PC, the total PC resistance contains three terms: $R_{sh}$, $R_{m\text{-}sample}$, and $R_{m\text{-}tip}$. Only the $R_{m\text{-}sample}$ is sensitive to the external magnetic field. As shown in Fig. 1b in main text, when perpendicular field is applied at 5 K, the ratio $\rho_{sample}(3T)/\rho_{sample}(0\,T)=41$. If we assume the ratio is the same for the local regime under the tip, it would have:

$$\frac{R_{m-sample}(3\,T)}{R_{m-sample}(0\,T)} = 41\ .$$

In Fig. 2a, we can obtain

$$R_{PC}(0\,T,\,6\,K) = 18.8\,\Omega;$$
$$R_{PC}(3\,T,\,6\,K) = 19.3\,\Omega.$$

One can estimate that

$$R_{m\text{-}sample} = 0.0125\,\Omega.$$

For Pt-Ir (80:20) tip, the resistivity is about $3 \times 10^{-8}$ $\Omega\cdot$m[6] at room temperature, which is much smaller than TaAs resistivity at room temperature (~$60 \times 10^{-8}$ $\Omega\cdot$m, see Fig. 1b in main text), so we can ignore the term $R_{m\text{-}tip}$, $R_m \approx R_{m\text{-}sample}$.

So the ratio

$$R_m/R_{sh} = 0.0125/18.8 = 0.066\%.$$

This means the contribution from Maxwell resistance for the total PC resistance is so little that the critical current cannot work in this condition.

For the other PC in our main text (Fig. 3b)

$$R_{PC}(0\,T,\,6\,K) = 4.2\,\Omega;$$
$$R_{PC}(3\,T,\,6\,K) = 4.4\,\Omega.$$

The calculating ratio $R_m/R_{sh} = 0.12\ \%$, so the critical current effect is also not applicable.

## IV Topological superconductivity protected by mirror symmetries

In this section we propose that the experimental observations can be interpreted with a minimal model of topological superconductivity obtained in TaAs. As stated in the main text, we consider for convenience the superconductivity induced on the four Weyl cones A-D in Fig. 4f of the main text, while the superconductivity on other Weyl cones can be obtained via C4 transformation or studied in the similar way. Note that to determine the superconducting phase, with the superconducting gap being a small energy scale relative to those of the Weyl semimetal, only the low energy physics of the Weyl cones are relevant. Thus, we can construct a minimal lattice model to study the present superconducting phase, as long as the lattice model gives the same Weyl cones A-D. The Hamiltonian of the normal state is given by

$$\mathcal{H}_0 = m_z\sigma_z + 2\mathcal{A}[3 - \cos k_0 \cos k_x - \cos k_y - \cos k_z]\sigma_z + 2\mathcal{B}\sin k_0 \cos k_x \sigma_x + 2\mathcal{B}\sin k_z \sigma_y$$
$$+ 2\mathcal{A}\sin k_0 \sin k_x \sigma_z s_y + 2\mathcal{B}\cos k_0 \sin k_x \sigma_x s_y, \quad (4)$$

where $\sigma_{x,y,z}$ are Pauli matrices acting on orbital space, $m_z$, $\mathcal{A}$, $\mathcal{B}$, and $k_0$ are system parameters determining the locations of the Weyl cones, and their details will not affect the results of the superconducting phase for our study. For convenience, we shall choose that $\mathcal{A} = \mathcal{B} = 1, m_z = -2$, and $k_0 = -1$. In this way, the four Weyl points are located at A $(1, -\pi/2, 0)$, B $(1, \pi/2, 0)$, C $(-1, \pi/2, 0)$, and D $(-1, -\pi/2, 0)$. It is trivial to know that the Hamiltonian respects the time-reversal symmetry $T = is_y K$, where $K$ is the complex conjugate operator. Moreover, it also preserves the mirror symmetries defined by
$$M_x = is_x, \ M_y = is_y. \tag{5}$$
The Hermitian parts are given by $M_x = s_x$, $M_y = s_y$. Since the superconductivity is induced on the (001) surface, which preserves $M_{x,y}$ symmetries, we seek for the nontrivial superconducting states respecting such mirror symmetries.

In the presence of superconductivity induced by tip, the Hamiltonian can be written in Nambu space, given by
$$\mathcal{H}_{BdG} = (\mathcal{H}_0 - \mu)\tau_z + \mathcal{H}_\Delta, \tag{6}$$
where $\mu$ is the chemical potential. The uniform superconducting pairing must occur between Weyl cones A (B) and C (D), since other pairing orders have nonzero center-of-mass momenta and are modulated in position space, which in general costs more energy. To keep the mirror symmetries, one can verify that the possible pairing terms of the superconducting order which can fully gap out the bulk include: (spin singlet) $\mathcal{H}_{\Delta_s} = \Delta_s s_y \tau_y$; (spin triplet) $\mathcal{H}_{\Delta_{px}} = \Delta_{px}(k_x)s_y\tau_x$ and $\mathcal{H}_{\Delta_{py}} = \Delta_{py}(k_y)\tau_y$, with $\Delta_{pi}(-k_i) = -\Delta_{pi}(k_i)$. Note that in the presence of superconductivity, the mirror symmetry $M_y$ has a second choice as $M_y = s_y\tau_z$. The mirror invariant planes for the mirror symmetry $M_i$ are $k_i = 0, \pi$, with $i = x, y$. The 3D mirror topological state can be characterized by the mirror topological invariant defined by[7]
$$N_{M_i} = sgn[C_1(k_i = 0) - C_1(k_i = \pi)][|C_1(k_i = 0)| - |C_1(k_i = \pi)|]. \tag{7}$$
The mirror Chern numbers $C_1(k_i = 0, \pi)$ are calculated on the 2D mirror planes $k_i = 0, \pi$, which define the 2D subsystems characterized by the Hamiltonian
$$\mathcal{H}_{2D}^{k_i=0,\pi} = \mathcal{H}_{BdG}(k_i = 0, \pi; \ k_\perp)P_+^i, \tag{8}$$
with $P_+^i$ being the operator projecting the system onto the subspace with mirror eigenvalue $M_i = +1$ and $k_\perp$ the 2D momentum perpendicular to $k_i$. Using the eqs. (5-8) one can calculate the mirror Chern number directly and then the mirror topological invariant. It is easy to show that for any of the above pairing orders, the mirror Chern number vanishes for the planes $k_x = 0, \pi$, i.e.
$$C_1(k_x = 0) = C_1(k_x = \pi) = 0.$$
Thus $N_{M_x} = 0$. For the mirror operator $M_y$, the invariant $N_{M_y}$ is shown to be zero as well for the pairing orders $\Delta_s$, and $\Delta_{px}$. However, as we show below, for the pairing order $\Delta_{py}$, the mirror topological invariant $N_{M_y}$ is nonzero.

We note that the results are independent of $k_0$. Then, for simplicity, we take $k_0 = 0$. In the eigenspace of $M_y = \pm 1$, we can rewrite the Hamiltonian as
$$\mathcal{H}_{BdG} = \begin{bmatrix} \mathcal{H}_+ + \Delta_{py}(k_y)\tau_y & 0 \\ 0 & \mathcal{H}_- + \Delta_{py}(k_y)\tau_y \end{bmatrix}, \tag{9}$$
where
$$\mathcal{H}_\pm = m_z\sigma_z\tau_z + 2\mathcal{A}[3 - \cos k_x - \cos k_y - \cos k_z]\sigma_z\tau_z + 2\mathcal{B}\sin k_z \sigma_y \tau_z \pm 2\mathcal{B}\sin k_x \sigma_x \tau_z.$$
It is important that $\Delta_{py}(k_y) = 0$ at the mirror invariant planes. Then for the subspace $\mathcal{H}_+$, the electron sector gives the mirror Chern number $C_1(k_y = 0) = 1$, and $C_1(k_y = \pi) = 0$, giving $N_{M_y} = 1$. The contribution by the hole sector is the same. Thus the total mirror topological invariant $N_{M_y} = 2$.